\documentclass[12pt]{article}

\usepackage{amssymb}
\usepackage{amsmath}
\usepackage{times}
\usepackage{graphicx}
\usepackage[all]{xy}

\newcommand{\vs}[1]{\vspace*{#1}}

\newcommand{\nn}{\nonumber}

\newcommand{\wt}{\widetilde}
\newcommand{\half}{\frac12}
\newcommand{\halfi}{\frac{i}{2}}
\newcommand{\beq}{\begin{equation}}
\newcommand{\eeq}{\end{equation}}
\newcommand{\beqa}{\begin{eqnarray}}
\newcommand{\eeqa}{\end{eqnarray}}
\newcommand{\bseq}{\begin{subequations}}
\newcommand{\eseq}{\end{subequations}}

\newcommand{\fr}{\frac}
\newcommand{\bg}{\textbf{g}}
\newcommand{\mn}{{\mu \nu}}

\newcommand{\bra}[1]{\langle #1|}
\newcommand{\ket}[1]{|#1\rangle}

\newcommand{\phiqn}{\ket{q_1,\cdots, q_n}}

\newcommand{\psiqn}{\ket{q_1,\cdots, q_n}_\star}

\newcommand{\cA}{\mathcal{A} }
\newcommand{\cE}{\mathcal{E}}
\newcommand{\cF}{\mathcal{F}}
\newcommand{\cH}{\mathcal{H}}
\newcommand{\cL}{\mathcal{L}}

\newcommand{\cU}{\mathcal{U}}

\newcommand{\phase}{\cE(q_1,\cdots,q_n)}

\newcommand{\poin}{Poincar\'{e} }

\newcommand{\bk}{\textbf{k}}
\newcommand{\bq}{\textbf{q}}
\newcommand{\bp}{\textbf{p}}

\newcommand{\hp}{\hat{p}}
\newcommand{\hq}{\hat{q}}
\newcommand{\Sskp}{S_\star^{\pm s}(\bq, p)}
\newcommand{\SsLkp}{S_\star^{\pm s}(\Lambda\bq, \Lambda p)}
\newcommand{\lphase}{\text{exp}\left[\pm i s\Theta(\bq,\Lambda)\right]}

\newcommand{\epms}{\epsilon_{\pm}^{~\mu_1*}(\hq)\cdots \epsilon_{\pm}^{~\mu_s*}(\hq)}
\newcommand{\Msl}{M_\star^{\pm\mu_1\cdots\mu_s}(\bq,p)}

\begin{document}

\nopagebreak

\baselineskip=16pt

\begin{titlepage}

\begin{center}
\vspace*{2cm}

{\Large\bf On Charge Conservation
and The Equivalence Principle \\
in the Noncommutative Spacetime}
\vspace{1cm}

Youngone Lee
\vspace{5mm}

{\it Center for Quantum Spacetime, Sogang University,
\\
 Seoul 121-742, Korea.}
\vspace{1cm}

\end{center}

\begin{abstract}
 We investigate one of the consequences of the twisted Poincar\'{e} symmetry.
We derive the charge conservation law and show that  the equivalence
principle is satisfied in the canonical noncommutative spacetime. We
applied the twisted Poincar\'{e} symmetry to the Weinberg's analysis
\cite{Weinberg}. To this end, we generalize our earlier construction
of the twisted S matrix \cite{Bu},
which apply the noncommutativity to the fourier modes,
to the massless fields of integer spins.
 The transformation formula for the twisted S matrix for the massless
fields of integer spin has been obtained. For massless fields of
spin 1, we obtain the conservation of charge, and the universality
of coupling constant for massless fields of spin 2, which can be
interpreted as the equality of gravitational mass and inertial mass,
i.e., the equivalence principle.
\end{abstract}


\footnotetext{youngone@yonsei.kr}
\setcounter{footnote}{0}
\end{titlepage}

\pagebreak

\section{Introduction}

In effort to construct an effective theory of quantum gravity at the
Planck scale, noncommutativity of the spacetime has been considered.
The canonical noncommutative spacetime has the commutation relations
between the coordinates \cite{Doplicher},
\begin{eqnarray}
\label{nc1} \left[x^\mu, x^\nu\right] &=& i \theta^{\mu\nu},
\end{eqnarray}
where $\theta^{\mn}(\mu,\nu = 0 \mbox{ to } 3)$ is a constant
antisymmetric matrix.

Field theories in the canonical noncommutative spacetime can be
replaced by field theories in the commutative spacetime with the
Moyal product (The Weyl-Moyal correspondence \cite{Weyl}). One of
the significant problems of those theories is that they violate the
Lorentz symmetry. One finds that the symmetry group is
$SO(1,1)\times SO(2)$ instead of the Lorentz group, $SO(1,3)$. Since
there is no spinor or vector representations in that symmetry group,
most of the earlier studies performed by using the spinor, vector
representations of the Lorentz group can not be justfied. Moreover,
the factors $1(-1)$ are multiplied for a boson(fermion)loop without
knowing the spin-statistics relation.

To get around this, Chaichian et.al.\footnote{
Oeckl \cite{Oeckl}, Wess \cite{Wess} have proposed
the same deformed \poin algebra.
} have deformed the \poin
symmetry as well as its module space to which the symmetry acts
\cite{chaichian}. The twisted symmetry group has the same
representations as the original \poin group and at the same time
they successfully retain the physical information of the canonical
noncommutativity. The main idea was that one can change a classical
symmetry group to a quantum group, $ISO_\theta(1,3)$ in this case,
and twist-deform the module algebra consistently to reproduce the
noncommutativity. In their approach, the noncommutative parameter
$\theta^{\mn}$ transform as an invariant tensor. This reminds us the
situation that Einstein had to change the symmetry group of the
spacetime and its module space(to the Minkowski spacetime) when the
speed of light is required to be constant for any observer in an
inertial frame. Similarly, Chaichian et.al. have required the change
of the Hopf algebra with its module algebra so that any observer in
an inertial frame feel the noncommutativity in the same way. For the
$\kappa$-deformed noncommutativity, Majid and Ruegg found the
$\kappa$-deformed spacetime  \cite{MajidRuegg} as a module space of
the $\kappa$-deformed symmetry after Lukierski et.al. discovered the
symmetry \cite{Lukierski}. The real benefit of the twist is in the
use of the same irreducible representations of original theories
unlike general deformed theories, as in the case of the
$\kappa$-deformed theory.

Recently, groups of physicists have constructed the quantum field
theory in the noncommutative spacetime by twisting the quantum space
as a module space  \cite{chaichianprl},\cite{Bala},\cite{Bu}.
Espcially, Bu et.al.  have proposed a twisted S-matrix as well as a
twisted Fock space for consistency \cite{Bu}. There we have obtained the
twisted algebra of the creation and annihilation operators and the
spin-statistics relation by applying the twisted \poin symmetry on
the quantum space consistently. The analysis of this paper is mainly
based on this work.

These works justify the use of the irreducible representations of
the \poin group and the sign factors being used in the earlier
studies. Are these works merely a change of viewpoint?
Mathematically they look equivalent and seem to have equal amounts
of information. But when the physics is concerned the action of
symmetry becomes more subtle than it seems because it confines
possible configurations of physical systems. In this article, we
present an example showing the role of the twisted symmetry for
solving physics problems, especially in the canonical noncommutative
spacetime. As an example, we derive the conservation law of charge
and show the fact that the equivalence principle is satisfied even
in the noncommutative spacetime. In this derivation we consider spin
1 and 2 massless fields for the photon and the graviton,
respectively. For this purpose, we extend our earlier study on the
scalar field theory to more general field theories and investigate a
noncommutative version of Weinberg's analysis
\cite{Weinberg,Weinberg1,Weinberg2}.

Actually, there are many studies for the relation between
noncommutativity and gauge theory \cite{Bonora}, \cite{WessGa},
\cite{Chaichiangauge}, \cite{Fatoll} or between the noncommutativity
and the gravity \cite{Rivelles},\cite{Banados},\cite{WessGr}
and between them \cite{Yang}. And
there were many argument whether the equivalence principle is
satisfied at the quantum level. Some people argued that the
equivalence principle is violated in quantum regime
\cite{Ellis},\cite{Adunas}, while there are studies which show
non-violation of the equivalence principle \cite{Obukhov}. Whether
the principle of equivalence is violated or not is an important
issue for quantum gravity because the principle is the core of the
general relativity.

The paper is organized as follows. We extend our previous
construction of the $S_\star$ matrix to the massless fields of
integer spin
 after giving a brief review on the construction and the properties of the $S_\star$ matrix.
We give an exact transformation formula for the $S_\star$ matrix
elements in section \ref{Smatrix}. We give the consequence of
requiring the twist invariance to the $S_\star$ matrix elements for
the scattering process. These results lead to the charge
conservation law for the spin 1 field theory and the universality of
the coupling constant for spin 2 field in noncommutative spacetime
in section \ref{NCEQ}. Finally, we discuss the implication of the
twisted symmetry and its applicability to other issues in section
\ref{discuss}. We give some related calculations of the polarization
vector, the noncommutative definition of the invariant $M$ function,
and a twisted transformation formula for the $S_\star$ matrix in the
Appendices.

\section{Properties of the general $S_\star$ matrix}
\label{Smatrix}

\subsection{A short introduction of useful properties of the twist-deformation}
\label{Introtwist}

An algebra with a product $\cdot$ and a coalgebra with a coproduct
$\Delta$ constitute a Hopf algebra if it has an invertible element
$S$ called antipode and with some compatibility relations.
For a Lie algebra $\bg$, 
there is a unique universal enveloping algebra $\cU(\bg)$ 
which preserves the Lie algebra properties
in terms of unital associative algebra. 
The Hopf algebra of a Lie algebra $\bg$ is denoted as
$\cH\equiv\{\cU(\bg),\cdot,\Delta,\epsilon,S\}$, where $\cU(\bg)$ is
an universal enveloping algebra of the corresponding algebra $\bg$
and we denotes the counit as $\epsilon$. The Sweedler notation is
being widely used for a shorthand notation of the coproduct, $\Delta
Y =\sum Y_{(1)}\otimes Y_{(2)}$ \cite{Majid}.

The action of a Hopf algebra $\cH$ to a module algebra $\cA$ is
defined as
\begin{equation}
Y\rhd (a \cdot b) = \sum (Y_{(1)}\rhd a) \cdot (Y_{(2)}\rhd b),
\label{action}
\end{equation}
where $a,b \in \cA$, the symbol $\cdot$ is a multiplication in the
module algebra $\cA$, and the symbol $\rhd$ denotes an action of the
Lie generators $Y\in U(\bg)$ on the module algebra $\cA$. The
product $\cdot$ in $\cH$ and the multiplication $\cdot$ in $\cA$
should be distinguished.

If there is an invertible 'twist element', $\cF=\sum
\cF_{(1)}\otimes \cF_{(2)} \in \cH\otimes\cH$, which satisfies
\begin{eqnarray}
\label{2cocycle} (\mathcal{F}\otimes 1)\cdot (\Delta\otimes
\mbox{id})\mathcal{F}
&=&(1\otimes \mathcal{F})\cdot  (\mbox{id}\otimes \Delta)\cF,\\
(\epsilon \otimes \mbox{id}) \cF=&1&=(\mbox{id}\otimes \epsilon)\cF
\label{counital},
\end{eqnarray}
one can obtain a new Hopf algebra
$\cH_\cF\equiv\{\cU_\cF(\bg),\cdot,\Delta_\cF,\epsilon_\cF,S_\cF\}$
from the original one. The relations between them are
\begin{eqnarray}
\Delta_\cF Y = \cF \cdot \Delta Y \cdot \cF^{-1} &,&
\epsilon_\cF (Y) = \epsilon (Y), \nn\\
S_\cF (Y) =  u \cdot S(Y)\cdot u^{-1} &,&u = \sum \cF_{(1)}\cdot
S(\cF_{(2)}), \label{tHopf}
\end{eqnarray}
with the same product in the algebra sector. The 'covariant'
multiplication of the module algebra $\cA_\cF$ for the twisted Hopf
algebra $\cH_\cF$ which maintain the form of Eq.(\ref{action}) is
given as \beqa \label{smul} (a\star b)=\cdot[\cF^{-1}~(a\otimes b)].
\eeqa

From the above relations, one can derive an important property of
the twist such that it does not change the representations of the
algebra: \beqa D_\cF(Y)(a\star b)
&=&\star~[\Delta_\cF Y~(a\otimes b)]\nn\\
&=&\cdot~[\cF^{-1}\cdot
\cF\Delta_0 Y\cF^{-1}(a\otimes b)]\nn\\
&=&\cdot~[\Delta_0 Y~
\cF^{-1}(a\otimes b)]\nn\\
&=&D_0(Y)(a\star b), \eeqa where representations of the coproduct
and the twist element is implied, i.e., \beqa D[\Delta Y]=\sum
D(Y_{(1)})\otimes D(Y_{(2)}),~~~~ D[\cF]=\sum D(\cF_{(1)})\otimes
D(\cF_{(2)}). \eeqa The above considerations lead us to the golden
rule: \textit{The irreducible representations are not changed by a
twist and one can regard the covariant action of a twisted Hopf
algebra on a twisted module algebra as the action of the original
algebra on the twisted module algebra.}

\subsection{The $S_\star$ matrix and its twist invariance}
\label{review}

Recently, a quantum field theory has been constructed in such a way
to preserve the twisted \poin symmetry \cite{Bu}. There we confined the
construction for the space-space noncommutativity. It is hard to
know whether one can construct a consistent twist \poin invariant
field theory satisfying the causality in the case of space-time
noncommutativity. They have tried to apply the twisted symmetry to
quantum spaces consistently, especially to the algebra of the
creation and annihilation operators ($a^\dagger_{ p}$ and $a_p$). As
a result, they obtained the twisted algebra of quantum operators. If
we use a shorthand notation,  $p\wedge q = p_\mu \theta^{\mn}q_\nu$,
the twisted algebra of $a^\dagger_{ p}$ and $a_p$ can be denoted as:
\beqa c_p\star~c_q = e^{-\halfi\wt{p}\wedge\wt{q}} ~c_p\cdot~c_q,
\eeqa where $c_p$ can be $a_p$ or $a^\dagger_{p}$,
$\wt{p}~\equiv-p(p)$ for $c_p = a_p(a^\dagger_{p})$ and $\cdot$
denotes the ordinary multiplication of operators in the commutative
theories.

This twisted algebra naturally leads to the twisted form of Fock
space, S-matrix and quantities related to the creation and
annihilation operators. Thus, we obtain the twisted basis of Fock
space and $S_\star$-matrix: \beqa \phiqn \rightarrow \psiqn
=\phase\phiqn, \eeqa where $\phase =\exp{\biggl(-\halfi\sum_{i<j}^n
q_i\wedge q_j \biggr)}$ is a phase factor which has the interesting
properties \cite{Bu}, and \beqa \label{tsmatrix} S \rightarrow
S_\star = \sum_{k=0}^{\infty}\fr{(-i)^k}{k!} \int d^4x_1\cdots
d^4x_k~ T\left\{\cH_I^\star(x_1)\star\cdots\star
\cH_I^\star(x_k)\right\}, \eeqa where $T$ denotes the time ordering
and $\cH_I^\star(x)$ is an interaction Hamiltonian density in the
Dyson formalism.

The explicit form of the $S_\star$ matrix elements for the scalar
$\phi^n$ theoryin the momentum space is:
\beqa \label{Sfilk}
_{\star}\bra{\beta}S_\star\ket{\alpha}_\star
= \cE(-\beta,\alpha)
\sum_{k=0}^{\infty}(-ig)^k \int_{Q_1}\cdots\int_{Q_k}
\sum_{c_{Q_1}\cdots c_{Q_k}} \cE(\wt{Q}_1)\cdots\cE(\wt{Q}_k)
\bra{\beta}S^k(\wt{Q}_1\cdots,\wt{Q}_k) \ket{\alpha},
\eeqa where
$\wt{Q}$ is the shorthand notation for $(\wt{q_1},\ldots,\wt{q_n})$
\cite{Bu}.

In the above, $\bra{\beta}S^k(\wt{Q}_1\cdots,\wt{Q}_k) \ket{\alpha}$
is a $g^k$ order term of the  S-matrix element of the commutative
theory where $g$ is the coupling constant of the theory. From the
momentum conservation, i.e., delta functions in the
$\bra{\beta}S^k(\wt{Q}_1\cdots,\wt{Q}_k) \ket{\alpha}$, one can show
that the $S_\star$ matrix element,
$_{\star}\bra{\beta}S_\star\ket{\alpha}_\star$, can be represented
by Feynman diagrams with extra phase factors $\cE(\wt{Q})$ for each
vertex. The phase factors drastically change the predictions of the
theory. This result agrees with Filk's result\cite{Filk}, but we
have overall factors $\cE(-\beta,\alpha)$ corresponding to external
lines in the Feynman diagram which originated from the twisted Fock
space. From the above considerations, the new modified Feynman
diagrams can be obtained from the untwisted ones by changing the
phase factors from 1 to $\cE(\wt{Q}_i)$ at each vertex.

The twist invariance of this prescription of the $S_\star$ matrix is
not manifest because non-locality of the interactions may violate
the twist invariance of the $S_\star$ matrix, in general, i.e.,
\beqa \left[\cH_I^\star(x),\cH_I^\star(y)\right]_\star \neq 0
~\mbox{ for spacelike}~(x-y). \eeqa However, we see from the form of
$S_\star$ matrix in Eq.(\ref{Sfilk}), that the proposed $S_\star$
matrix is clearly twist invariant since it is constructed from phase
factors which are twist invariant, and the Feynman propagators.
Twisted product of fields operators satisfy \beqa
\bra{0}\psi(x)\star \psi(y)\ket{0} =\bra{0}\psi(y)\star
\psi(x)\ket{0}, \text{  for spacelike $(x-y)$.} \eeqa Hence we see
that the  Feynman propagator, same as the twist Feynman propagator
$\bra{0}T[\psi(x)\star\psi(y)]\ket{0}$, is twist invariant. From
this, the invariance of the $S_\star$ matrix elements follows
immediately.

\subsection{Generalization to arbitrary fields}
\label{extention}

We need to get the $S_\star$ matrix for massless field theories of
integer spin for the analysis of this paper. In the previous work
\cite{Bu}, we have constructed the $S_\star$ matrix for scalar field
theory and we have expected that the same formulation could be
possible for general field theories. In this section, we generalize
our argument used in that paper to obtain the form of the $S_\star$
matrix elements for massless field theories with spin 1 and 2. In
the analysis of this paper, we use the $(s/2,s/2)$ representation
for massless integer spin fields. The reason we use it and the
considerations for the other representations are given in section
\ref{discuss}.

We have used the perturbation theory in our formulation of the
$S_\star$ matrix in Eq.(\ref{tsmatrix}). Another assumption was the
particle interpretation. That is, the field operators are
represented as linear combinations of the creation and annihilation
operators and the fields of spin $s$ transform(twist) as\footnote{
Recently, Joung and Mourad \cite{Joung} have argued that a covariant
field linear in creation and annihilation operators does not exist.
But the definition of creation and annihilation operators in that
paper is different from ours. We try to follow the view that we use
the same fundamental quantities as of those in the untwisted
theories changing the algebras only. The field operators in linear
combinations of the creation and annihilation operators can be
justified in this basis, because the free fields are the same as the
commutative case.} : \beqa \label{ftlorentz}
[D^s_\theta(\Lambda^{-1})]^A_{~B} ~\hat{\Psi}_\theta^{~B}(\Lambda
x+a) =U_\theta(\Lambda,a)~
\hat{\Psi}_\theta^{~A}(x)~U^{-1}_\theta(\Lambda,a), \eeqa where
$D^s$ denotes the irreducible representation for spin $s$. Since
translations act homogenously on the fields, twisted tensor fields
can be written as \beqa \hat{\Psi}_\theta^{~A}(x)
=\int_p\sum_\sigma[a_\sigma(p)~\epsilon_\theta^{~A}(p,\sigma)~e^{ip\cdot
x} +b^\dagger_\sigma(p)~\chi_\theta^{~A}(p,\sigma) ~e^{-ip\cdot x}],
\eeqa where $A$ denotes the tensor index, $A\equiv (a_1\cdots a_s)$
for massless spin $s$ fields and  $\sigma$ for the helicity indices.
Thus, the transformation relation of the fields,
Eq.(\ref{ftlorentz}),  can be reduced to \beqa \label{ftnc}
[D^s_\theta(\Lambda^{-1})]^A_{~B} ~\hat{\Psi}_\theta^{~B}(\Lambda x)
=U_\theta(\Lambda)~
\hat{\Psi}_\theta^{~A}(x)~U^{-1}_\theta(\Lambda). \eeqa In order
that the field operators transform correctly, the 'polarization'
tensor $\epsilon_\theta^{~A}(p,\sigma)$ are to be transformed as
\beqa \epsilon_\theta^{~A}(p,\sigma) &=&
\fr{1}{(2\pi)^3\sqrt{2\omega_p}}
~[D_\theta^s(\cL(p))]^A_{~B}~\epsilon_\theta^{~B}(k,\sigma),\nn\\
\chi_\theta^{~A}(p,\sigma) &=& \fr{1}{(2\pi)^3\sqrt{2\omega_p}}
~[D_\theta^s(\cL(p))]^A_{~B}~\chi_\theta^{~B}(k,\sigma')
~C_{\sigma'\sigma}^{-1}, \eeqa where $\cL(p)$ is the Lorentz
transformation\footnote{ $\cL(p)$ is a Lorentz transformation that
takes the standard momentum $k^\mu$ to
$p^\mu\equiv(|\bold{p}|,\bold{p})$ for massless fields. For the
massless case $\cL(p)$ satisfies $\cL(p)=R(\hat{p})B(|\bold{p}|)$
where $R(\hat{p})$ is a rotation which takes the direction of the
standard momentum $\bk$ to the direction of $\bold{p}$ and
$B(|\bold{p}|)$ is the boost along the $\bold{p}$ direction
\cite{Weinberg}. } which reflects the little group  and $C$ is a
matrix
 with the properties \cite{Weinberg1},
\beqa C^*~C =(-1)^{2s},~~~C^\dagger~C=1. \eeqa

Since the most important properties of the twist is that the
representations are the same as in the original group, the \poin
group in this case, the above $D^s_\theta(\Lambda^{-1})$ and
$D_\theta^s(\cL)$ can be replaced by $D^s(\Lambda^{-1})$ and
$D^s(\cL)$, respectively. Thus the $\theta$-dependance remains only
in the polarization tensors. Furthermore, one can see that there's
no $\theta$-dependance in $\epsilon_\theta^{~A}(\sigma)$ and
$\chi_\theta^{~A}(\sigma')$. We can show this by considering the
case of $\epsilon_\theta^{~A}(\sigma)$. From the group property of
the twisted Lorentz transformations, the related transformation to
the polarization tensor $\epsilon_\theta^{~A}(\sigma)$ can be
written as: \beqa [D^s_\theta(\Lambda^{-1})]^A_{~B}
~[D_\theta^s(\cL)]^C_{~D}~\epsilon_\theta^{~D}(\sigma)
=[D^s_\theta(\Lambda^{-1}\cdot \cL)]^A_{~B}
~\epsilon_\theta^{~B}(\sigma). \eeqa In order for  this
transformation to be a twist transformation, it is to be satisfied
\beqa [D^s_\theta(\Lambda^{-1}\cdot \cL)]^A_{~B}
~\epsilon_\theta^{~B}(\sigma) = [D^s(\Lambda^{-1}\cdot \cL)]^A_{~B}
~\epsilon_\theta^{~B}(\sigma), \eeqa i.e., twisted transformation
has the same representation as the untwisted one. From the primary
relation between the twist and the module algebra, one can see that
the form of $\epsilon_\theta^{~A}(\sigma)$, twisted version of the
commutative polarization tensor, should be \beqa
\epsilon_\theta^{~a_1\cdots a_n}(\sigma) =~\cdot
[\cF_n^{-1}\triangleright \epsilon^{a_1}(\sigma)\otimes\cdots\otimes
\epsilon^{a_n}(\sigma)], \eeqa where $\cF_n^{-1}$ can be obtain
from\footnote{ The associativity of the twisted product $*$ guaranty
that it can be obtained by successive applications of
Eq.(\ref{smul}). } the $\cF_\theta$, and $\cF_\theta$ is the very
twist element corresponding to the canonical noncommutativity: \beqa
\label{Fth} \cF_\theta = \exp\left(\halfi \theta^{\alpha\beta}
P_\alpha\otimes P_\beta\right). \eeqa Since
$P_\alpha\triangleright\epsilon^{a}(\sigma)=0$, we obtain \beqa
\label{poltensor} \epsilon_\theta^{~a_1 a_2\cdots a_s}(\sigma)
\equiv \epsilon^{~a_1 a_2\cdots a_s}(\sigma)
=\epsilon^{(a_1}_\sigma\epsilon^{a_2}_\sigma
\cdots\epsilon^{a_s)}_\sigma, \eeqa where $\epsilon^{(a_1 a_2\cdots
a_s)}(\sigma)$ is the polarization tensor in the corresponding
commutative field theory. This relation leads to the relation \beqa
\hat{\Psi}_\theta^{~A}(x) \equiv\hat{\Psi}^{A}(x). \eeqa This
relation is expected because the twist does not change the
representations and we write the field operators by using an
irreducible representations of the symmetry group. Explicit
calculations for massless $(s/2,s/2)$ tensor representation which
shows the non-dependance on the $\theta$ is given in the Appendix
\ref{apa}.

Therefore, one can safely use the same representations for the field
operators in the twisted theory as those of the untwisted ones. What
really be twisted are the multiplications of the creation and
annihilation operators only. Since the multiplication of the
creation and annihilation operators between different species of
particles act like composite mappings on the Hilbert space, one can
construct a module algebra from them. That is, when $c_p$'s and
$d_p$'s are the creation and annihilation operators corresponding to
different species, their twisted multiplications are \beqa
c_p\star~d_q = e^{-\halfi\wt{p}\wedge\wt{q}} ~c_p\cdot~d_q, \eeqa
where $c_p(d_p)$ can be $a_p(b_p)$ or
$a^\dagger_{p}(b^\dagger_{p})$, and $\wt{p}$ and $\wt{q}$ are
defined as in section \ref{review}.

Consequently, the scheme for twisting $S$-matrix used in \cite{Bu}
applies for general field theories. Twist invariance of the
$S_\star$ matrix for general field theories follows immediately. As
in the scalar field theory, the amplitudes can be obtained by
multiplying the phase factor $\phase$ for each vertex in the Feynman
diagram of the untwisted theories.

\subsection{Exact transformation formula for the $S_\star$ matrix element for massless fields}

Transformation formula for the $S_\star$ matrix element
corresponding to the process in which a massless particle is emitted
with momentum $\bq$ and helicity $s$ can be inferred as (Appendix
\ref{apc}) \beqa \label{Stk} \Sskp= \sqrt{ \fr{|\Lambda \bq|}{|\bq|}
}\lphase\SsLkp. \eeqa The $S_\star$ matrix can be written as the
scalar product of a polarization tensor and the $M_\star$
function(Appendix \ref{apc}):
\beqa \label{stran}
\Sskp=
\fr{1}{\sqrt{2|\bq|}}\epms (M_\star^\pm)_{\mu_1\cdots\mu_s}
(\bq,p),
\eeqa
where the $M_\star$ function twist-transform covariantly as
\beqa
\Msl=\Lambda_{\nu_1}^{~\mu_1}\cdots\Lambda_{\nu_s}^{~\mu_j}
M_\star^{\pm\nu_1\cdots\nu_s}(\Lambda\bq,\Lambda p). \eeqa

The form of the $S_\star$ matrix element in Eq.(\ref{stran}) appears
to break the twisted \poin symmetry because the polarization vectors
do not satisfy the Lorentz covariance, rather they satisfy \beqa
\label{tepsilon}
(\Lambda_\nu^{~\mu}-q^\mu\Lambda_\nu^{~0}/|\bq|)\epsilon_\pm^{~\nu}(\Lambda
q) =\text{exp}\left\{\pm i \Theta(\bq,\Lambda)\right\}
\epsilon_\pm^{~\mu}(\hq). \eeqa Hence requiring the twist invariance
of the $S_\star$ matrix would lead to a constraint relation between
the momentum, the polarization vectors and the $M_\star$ function.
From Eq.(\ref{Stk}) and Eq.(\ref{tepsilon}), the $S_\star$ matrix
element in Eq.(\ref{stran}) can be written as \beqa
\Sskp&=&\fr{1}{\sqrt{2\bq}}\lphase [\epsilon_\pm^{~\mu_1}(\Lambda q)
-(\Lambda q)^{\mu_1}\Lambda_{\nu}^{~0}\epsilon_\pm^{\nu}
(\Lambda q)/{|\bq|}]^*\nn\\
&&~~~~~~~\cdots [\epsilon_\pm^{~\mu_s}(\Lambda q) -(\Lambda
q)^{\mu_s}\Lambda_{\nu}^{~0}\epsilon_\pm^{\nu} (\Lambda
q)/{|\bq|}]^* (M_\star)_{\pm\mu_1\cdots\mu_s}(\Lambda \bq,\Lambda
p). \eeqa Requiring the twist invariance of the $S_\star$ matrix
element results in: \beqa q^{\mu_1}\epsilon_\pm^{~\mu_2*}
(\hq)\cdots\epsilon_\pm^{~\mu_s*}(\hq)
{M_\star^\pm}_{\mu_1\cdots\mu_s}=0. \eeqa This leads us to the
desired identities: \beqa \label{Ward} q_\rho
M_\star^{\pm\rho\mu_2\cdots\mu_s}(\bq,p)=0. \eeqa

\section{Charge conservation and the equivalence principle }
\label{NCEQ}

Since the analysis of the conservation law is just a noncommutative
generalization of Weinberg's work \cite{Weinberg}, the derivation of
this section will be fairly straightforward. As we saw in section
\ref{Smatrix}, the differences between the noncommutative field
theory and the commutative one are the phase factors at each vertex
of the Feynman diagrams.

\subsection{Dynamical definition of charge and gravitational mass}

We define the charge and the gravitational mass dynamically as the
coupling of the vertex amplitudes for the soft(very low energy)
photon and graviton, respectively. Consider the vertex amplitude for
the process that a soft\footnote{ This is to define the charge and
gravitational mass as a monopole, not as a multipole moments. }
massless particle of momentum $q$ and spin $s$ is emitted by a
particle of momentum $p$ and spin $J$. Since the only tensor which
can be used to form the invariant $M$ function is known to be
$p^{\mu_1}\cdots p^{\mu_s}$, the noncommutative $M$ function,
$M_\star$ function, is given by the commutative $M$ function with
phase factor $\cE(\wt{q},\wt{p},\wt{p}')$ multiplied at each vertex.
That is, the vertex amplitude can be written as \beqa \Gamma_\star~
\propto
 \fr{\cE(\wt{q},\wt{p},\wt{p}')}{2E(\bp)\sqrt{2|\bq|}}
~p_{\mu_1}\cdots p_{\mu_s}
~\epsilon_\pm^{~\mu_1*}(\hq)\cdots\epsilon_\pm^{~\mu_s*}(\hq). \eeqa
When the emitting particle has spin $J$ we have to multiply to the
vertex amplitude $\delta_{\sigma\sigma'}$ \cite{Weinberg}. Thus, the
explicit form of the vertex amplitudes can be written as: \beqa
\fr{2i(2\pi)^4\cdot \bold{e_s}\cdot \delta_{\sigma\sigma'}
[p_\mu~\epsilon_\pm^{~\mu*}(\hq)]^s }
{(2\pi)^{9/2}[2E(\bp)]\sqrt{2|\bq|}}
\cdot\cE(\wt{q},\wt{p},\wt{p}'), \eeqa where $\bold{e_s}$ is a
coupling constant for emitting a soft massless particle of spin $s$
(e.g., photon and graviton).

These coupling constants for emitting a soft particle can be
interpreted as: $\bold{e_1}\equiv e$ as the electric charge, and
$\bold{e_2}\equiv\sqrt{8\pi G}g$, with $g$ the ratio of the
gravitational mass and the inertial mass. Let us consider a near
forward scattering of the two particles $A$ and $B$ with the
coupling $\bold{e}^s_{A}$ and $\bold{e}^s_B$, respectively. From the
properties of the phase factors \cite{Bu}, the phase factor for this
scattering, $\cE(\wt{q},\wt{p}_A,\wt{p}_{A'})
\cdot\cE(-\wt{q},\wt{p}_B,\wt{p}_{B'})$, goes to: \beqa
\cE(\wt{q},\wt{p}_A,\wt{p}_{A'})
\cdot\cE(-\wt{q},\wt{p}_B,\wt{p}_{B'}) &=&
\cE(\wt{p}_A,\wt{p}_{A'},\wt{q})
\cdot\cE(-\wt{q},\wt{p}_B,\wt{p}_{B'})\nn\\
&=&\cE(\wt{p}_A,\wt{p}_{A'}) \cdot\cE(\wt{p}_B,\wt{p}_{B'}). \eeqa
However, in the forward scattering limit, the direction of the
particles does not change, i.e. $\bp_A\parallel \bp'_A$. For
space-space noncommutativity $p_A\wedge p'_A$ goes to zero, i.e.,
the phase factor goes to 1.

Thus, when the invariant momentum transfer $t=-(p'_A-p_A)^2$ goes to
zero, using the properties of the polarization vectors in the
Appendix \ref{apb}, the $S_\star$ matrix element can be shown to
approach the same form as in the corresponding commutative quantity
which is easily calculated in a well chosen\footnote{ The coordinate
system in which $q\cdot p_A = q\cdot p_B =0$ \cite{Weinberg}. }
coordinate system, \beqa \fr{ \delta_{\sigma_A\sigma_{B'}}
\delta_{\sigma_B\sigma_{B'}}}{4\pi^2 E_A E_B t} \left[e_A
e_{B}(p_A\cdot p_B) + 8\pi G~g_{A} g_{B}\left\{(p_A\cdot p_B)^2
-\half m_A^{~2}m_B^{~2}\right\}\right]. \eeqa This coincidence is
quite special in the sense that the $S_\star$ matrix elements are
quite different from the $S$ matrix elements in the commutative
theory when there is a momentum transfer. If particle $B$ is at
rest, this gives \beqa \fr{ \delta_{\sigma_A\sigma_{A'}}
\delta_{\sigma_B\sigma_{B'}}}{\pi t} \left[-\fr{e_{A} e_{B}}{4\pi}+
G\cdot g_A\left(2E_A-\fr{m_A^{~2}}{E_A}\right) \cdot g_B m_B\right].
\eeqa Hence, one can interpret the coupling constant $e_A$ as the
usual charge of the particle $A$. Moreover one can identify the
effective gravitational mass of $A$ as \beqa
(m_g)_A=g_A\left(2E_A-\fr{m_A^{~2}}{E_A}\right). \eeqa For
nonrelativistic limit the $g_A$ can be interpreted as a ratio of the
gravitational mass and the inertial mass, i.e., \beqa
(m_g)_A=g_A\cdot m_A,&&E_A\simeq m_A. \eeqa Consequently, if $g_A$
does not depend on the species of $A$, it suggests that the
equivalence principle holds.

\subsection{Conservation law}
Consider a $S$ matrix element $S(\alpha\rightarrow\beta)$ for some
reaction $\alpha\rightarrow\beta$, where the states $\alpha$ and
$\beta$ consist of various species of particles. The same reaction
with emitting a soft massless particle of spin $s$ (photon or
graviton for $s=1$ or $2$), momentum $\bq$ and helicity $\pm s$ can
occur. We denote the corresponding $S$ matrix element as $S^{\pm
s}(\bq,\alpha\rightarrow \beta)$. Each amplitude of this process
breaks the Lorentz symmetry because a massless fields of $(s/2,s/2)$
representation break the symmetry \cite{WeinbergB}. However, a real
physical reaction, to which a $S$-matrix element (the sums of each
amplitude) correspond, should be Lorentz invariant. By requiring
this condition, Weinberg could obtain the conservation relations. In
this section, we investigate the similar relations by requiring the
twist invariance of the $S_\star$ matrix elements.

Suppose that a soft particle is emitted by $i$th particle
($i=1,\cdots,n$). Then by the polology of the conventional field
theory, the $S$ matrix elements will have poles at $|\bq|= 0$ when
an extra soft massless particle is emitted by one of the external
lines: \beqa \fr{1}{(p_i+\eta_i \cdot q)^2+m_i^{~2}}
=\fr{\eta_i}{2p_i\cdot q}, \eeqa where $\eta_i=1(-1)$ for the
emission of a soft particle from an out(in) particle, respectively.
By utilizing the above relation, one can obtain the $S$ matrix
elements for the soft massless particles in the $|\bq|\rightarrow 0$
limit \cite{Weinberg}: \beqa \label{sss} S^{\pm
s}(\bq,\alpha\rightarrow \beta) = \fr{1}{(2\pi)^{3/2}\sqrt{2|\bq|}}
\left( \sum_i \eta_i \bold{e_s}_i \fr{[p_i\cdot
\epsilon_\pm^{~*}(\hq)]^s}{(p_i\cdot q)} \right)
S(\alpha\rightarrow\beta). \eeqa

In the noncommutative case, from the result of the section
(\ref{Smatrix}) and by using the above relation, one can deduce the
$S_\star$ matrix elements for the same process in the noncommutative
spacetime: \beqa \label{ncsss} S_\star^{~\pm
s}(\bq,\alpha\rightarrow \beta) &=&
\sum_i\cE_i(\wt{q},\alpha\rightarrow\beta) \cdot
S_i^\pm(\bq,\alpha\rightarrow \beta)
\nn\\
&=& \sum_i\cE(\wt{q},\wt{p_i},\wt{p_i}+\wt{q}) \cdot
\cE_I(\wt{p_1},\cdots,\wt{p_i}+\eta_i\wt{q},\cdots,\wt{p}_n) \cdot
S_i^\pm(\bq,\alpha\rightarrow \beta)
\nn\\
&=& \sum_i \fr{\cE(q,\wt{p_i})} {(2\pi)^{3/2}\sqrt{2|\bq|}}
\left(\eta_i \bold{e_s}_i \fr{[p_i\cdot
\epsilon_\pm^{~*}(\hq)]^s}{(p_i\cdot q)} \right) \nn\\
&&
\hspace{1cm}
\times~
\cE_I(\wt{p_1},\cdots,\wt{p_i}+\eta_i\wt{q},\cdots,\wt{p}_n) \cdot
S^\pm_i(\alpha\rightarrow\beta),
\eeqa
where $\cE_I^i=
\cE_I(\wt{p_1},\cdots,\wt{p_i}+\wt{q},\cdots,\wt{p}_n)$ are the
phase factors in the internal process\footnote{ The integrals over
the internal loop momenta have been suppressed because they do not
affect the final result since the change in the phase factor occurs
only in the external lines.}. Since all the $\cE_I^i$'s are the same
when $q\rightarrow 0$ ($\cE_I^i=\cE_I$), one obtains in that limit,
\beqa \label{ncsq} S_\star^{~\pm s}(\bq,\alpha\rightarrow \beta) =
\fr{1}{(2\pi)^{3/2}\sqrt{2|\bq|}} \left( \sum_i \eta_i \bold{e_s}_i
\fr{[p_i\cdot \epsilon_\pm^{~*}(\hq)]^s}{(p_i\cdot q)} \right)
S_\star(\alpha\rightarrow\beta), \eeqa where we used the relation
${S_\star}(\alpha\rightarrow\beta) =\cE_I\cdot
S(\alpha\rightarrow\beta)$. From the transformation properties of
the $S_\star$ matrix elements, Eq.(\ref{stran}), we obtain, \beqa
\label{ssm} S_\star^{~\pm s}(\bq,\alpha\rightarrow \beta)
&\rightarrow&\fr{1}{\sqrt{2|\bq|}}
\epsilon_\pm^{~\mu_1*}(\hq)\cdots\epsilon_\pm^{~\mu_s*}(\hq)
(M_\star^\pm)_{\mu_1\cdots \mu_s}(\bq,\alpha\rightarrow \beta).
\eeqa Identifying the $S_\star$ matrix elements in Eq.(\ref{ncsq})
and Eq.(\ref{ssm}) gives the invariant $M_\star$ functions for spin
1 and 2: \beqa M_\star^\mu(\bq,\alpha\rightarrow\beta) &=&
\fr{1}{(2\pi)^{3/2}}\left(\sum_i \fr{e_i \cdot\eta_i
p_i^{~\mu}}{(p_i\cdot q)} \right)
S_\star(\alpha\rightarrow\beta),\nn\\
M_\star^{\mu\nu}(\bq,\alpha\rightarrow\beta) &=& \fr{8\pi
G}{(2\pi)^{3/2}}\left(\sum_i \fr{g_i \cdot\eta_i p_i^{~\mu}
p_i^{~\nu}}{(p_i\cdot q)} \right) S_\star(\alpha\rightarrow\beta).
\eeqa Requiring the twist invariance to the $S_\star$ matrix
elements, Eq.(\ref{Ward}), gives: \beqa 0= q_\mu
M_\star^\mu(\bq,\alpha\rightarrow\beta) &\rightarrow& \sum_i \eta_i
e_i=0,
\nn\\
0= q_\mu M_\star^{\mu\nu}(\bq,\alpha\rightarrow\beta) &\rightarrow&
\sum_i g_i \cdot \eta_i p_i^{\nu}=0, \eeqa in general. Hence, one
obtains the charge conservation law for the spin 1 fields. For
$s=2$, in order to satisfy the two relations, $\sum \eta_i p_i = 0$
(4-momentum conservation) and $\sum_i g_i \cdot \eta_i p_i^{\nu}=0$,
$g_i$ should be constants (i.e. independence of the particle
species). The universality of this coupling constant as a ratio of
gravitational mass and inertial mass shows that the equivalence
principle is satisfied even in the noncommutative spacetime.

\section{Discussion}
\label{discuss}

We have found that the conservation of charge and the equivalence
principle are satisfied even in the canonical noncommutative
spacetime. The derivation was fairly straightforward, once we have
constructed the $S_\star$ matrix for general field theories. The
assumption was that the quanta of the gravitation is massless spin
2, massless spin 1 for photon.

We extended the construction of the $S_\star$ matrix to general
fields, especially for the massless fields of integer spin. The
twisted Feynman diagrams can be constructed by the same irreducible
representations as those in the untwisted theories with the same
rule except for the different phase factors at each vertex. Hence,
we can say that the same reasoning apply to the massive fields.

We use the $(s/2,s/2)$ representation for the field operators mainly
because one can obtain the condition, Eq.(\ref{Ward}), requiring the
twist Lorentz invariance. In this representation, since the
polarization vectors are not Lorentz four vectors, each amplitude of
emitting a real soft massless particle violate the symmetry. We have
used this property to show the conservation law in this paper.
Another representation for the fields, the $(s,0)\oplus (0,s)$
representation, which has the parity symmetry, can be made Lorentz
covariant. One realizes that due to these properties one cannot
derive the conservation law by the method used in this paper.

Charge conservation in the noncommutative spacetime is expected from
the gauge symmetry and the noncommutative Noether theorem. However,
it is not quite certain whether the equivalence principle is
satisfied in the noncommutative spacetime. Since the noncommutative
spacetime is not locally Minkowski nor the local symmetry group is
$SO(1,3)$, one can not guarantee if the principle is satisfied.
But in our twisted symmetry context,
one can expect that the equivalence principle is satisfied
because the algebra structure is the same as the conventional theory
though the coalgebra structure is different.
The applicability of the $S$ matrix theoretic proof of the equivalence
principle given in this paper is restrictive because the analysis is
perturbative in nature. We expect that the conclusion of this paper
is to be one of the stepping stone towards the further understanding
of the nature of the principle in quantum gravity.

This paper shows an example for the usefulness of the twisted
symmetry to derive the physically important relations in the
noncommutative spacetime. We think that the twist analysis is
adequate for the schematic approaches to the noncommutative physics,
while the other approaches are suitable for the explicit
calculations though they are equivalent. The future applications of
the twisted symmetry to the other issues in this direction of
approach are expected.

\section*{Acknowledgement}
I express my deep gratitude to Prof. J. H. Yee for his support.
I would like to thank Dr. Jake Lee for helpful discussions
and Dr. A. Tureanu for his comments on the draft. 
This work was supported in part by Korea
Science and Engineering Foundation Grant No. R01-2004-000-10526-0,
and by the Science Research Center Program of the Korea Science and
Engineering Foundation through the Center for Quantum
Spacetime(CQUeST) of Sogang University with grant number R11 - 2005
- 021.

\begin{appendix}

\section{Exact calculation of the polarization tensor for massless fields}
\label{apa}

The creation and annihilation operators of the massless field
transform under the Lorentz transformation as \cite{WeinbergB}
\beqa
\label{taad} a^\dagger(\Lambda p,\sigma)&=& e^{-i\sigma
\Theta[W(\Lambda,p)]}
~U(\Lambda)~a^\dagger(p,\sigma)~U^{-1}(\Lambda),\nn\\
a(\Lambda p,\sigma)&=& e^{+i\sigma \Theta[W(\Lambda,p)]}
~U(\Lambda)~a(p,\sigma)~U^{-1}(\Lambda),
\eeqa
up to phase factors,
where $W(\Lambda,p)=\cL^{-1}(p)\Lambda^{-1}\cL(\Lambda p)$ denotes
the Wigner rotation defined as in \cite{WeinbergB}
(Appendix \ref{apc}), and the
$\Theta[W(\Lambda,p)]$ is the related angle to which the little
group corresponds. Hereafter, we abbreviate $ \Theta[W(\Lambda,p)]$
as $ \Theta(\Lambda,p)$. It is to be noted that we use $U(\Lambda)$
instead of $U_\theta(\Lambda)$ in here. This follows immediately
from the properties of the twist  in section \ref{Introtwist}.

The field $\hat{\Psi}_\theta^{~A}$ transform as \beqa \label{ftapdx}
[D^s(\Lambda^{-1})]^A_{~B}~\hat{\Psi}_\theta^{~B}(\Lambda x) &=&
U(\Lambda)~ \hat{\Psi}_\theta^{~A}(x)~U^{-1}(\Lambda). \eeqa In
order to satisfy the two relations (\ref{taad}), (\ref{ftapdx}), the
polarization tensor should transform as \beqa \epsilon_\theta^{~A}(
p,\sigma) =e^{-i\sigma \Theta(\Lambda,p)}
~[D^s(\Lambda^{-1})]^A_{~B}~ \epsilon_\theta^{~B}(\Lambda p,\sigma).
\eeqa When $\Lambda = W$ and $p=k$, the above relation goes to \beqa
~[D^s(W)]^A_{~B}~\epsilon_\theta^{~B}(k,\sigma) =e^{-i\sigma
\Theta(W,p)} \epsilon_\theta^{~A}(k,\sigma), \eeqa where $k$ denotes
the standard momentum. The Wigner rotation can be written as
$W(\phi,\alpha,\beta)=R(\phi)T(\alpha,\beta)$ because the little
group is isomorphic to $ISO(2)$ for massless fields
\cite{WeinbergB}.

Suppose that the field transforms as $(m,n)$ representation of spin
$s~(m+n=s)$. When the $\phi,\alpha$ and $\beta$ are infinitesimal,
$D^s(W)$ can be written as \beqa D^s(W)\simeq 1-i\phi(M_3+N_3)
+(\alpha+i\beta)(M_1-iM_2)+(\alpha-i\beta)(N_1+iN_2). \eeqa For
$M_-\equiv M_1-iM_2$, $N_+\equiv M_1+iM_2$, and $\Theta \rightarrow
\phi$ we have: \beqa \label{eqpol}
(M_3+N_3)~\epsilon_\theta(k,\sigma)
&=&\sigma~\epsilon_\theta(k,\sigma),\nn\\
M_{-}~\epsilon_\theta(k,\sigma)
&=&0,\nn\\
N_{+}~\epsilon_\theta(k,\sigma) &=&0, \eeqa which gives \beqa
M_3~\epsilon_\theta(k,\sigma)
&=&-m~\epsilon_\theta(k,\sigma),\nn\\
N_3~\epsilon_\theta(k,\sigma) &=&+n~\epsilon_\theta(k,\sigma). \eeqa
The $\epsilon_\theta(k,\sigma)$ satisfys the same equation as
$\epsilon(k,\sigma)$. Since the highest or lowest weight corresponds
to a unique state, one obtains the same polarization tensor as a
solution, $\epsilon_\theta(k,\sigma)=\epsilon(k,\sigma)$.

\section{Properties of the polarization vectors}
\label{apb}

Here, we summarize the properties of the polarization vector. The
properties of the polarization tensors of other rank,
Eq.(\ref{poltensor}), follows from it.

Solving the (\ref{eqpol}) for $\sigma = \pm 1$ gives the explicit
form of the polarization vector for the standard momentum as \beqa
\epsilon_{\pm}^{~\mu}(k) \equiv \fr{1}{\sqrt{2}}(1,\pm i,0,0), \eeqa
where we made the conventional choice of the phase. The polarization
vector of momentum $p$ is defined as \beqa \epsilon_{\pm}^{~\mu}(p)
= [\cL(p)]^\mu_{~\nu} ~\epsilon_{\pm}^{~\nu}(k), \eeqa where
$\cL(p)$ is the Lorentz transformation which takes $k$ to $p$, i.e.,
$p^\mu =[\cL(p)]^\mu_{~\nu}~ k^\nu$. Then the well known properties
of the polarization vectors can be deduced \cite{Weinberg}: \beqa
p_\mu\epsilon_{\pm}^{~\mu}(\hp)&=&0,\\
\epsilon_{\pm\mu}^{~~*}(\hp)\epsilon_{\pm}^{~\mu}(\hp)&=&1,~~~~~~~~
\epsilon_{\pm\mu}(\hp)\epsilon_{\pm}^{~\mu}(\hp)=0,\\
\epsilon_{\pm}^{~\mu*}(\hp)&=&\epsilon_{\mp}^{~\mu}(\hp),~~~~~
\epsilon_{\pm}^{~~0}(\hp)=0,\\
\sum_{\pm}\epsilon_{\pm}^{~\mu}(\hp)\epsilon_{\pm}^{~\nu*}(\hp)&=&
\eta^{\mu\nu}+(\wt{p}^\mu p^\nu+\wt{p}^\nu p^\mu)/{2|\bp|^2}\equiv
\Pi^{\mu\nu}
,~\wt{p}\equiv(|\bp|,-\bp),\\
\sum_{\pm}\epsilon_{\pm}^{~\mu_1}(\hp)\epsilon_{\pm}^{~\mu_2}(\hp)
\epsilon_{\pm}^{~\nu_1*}(\hp)\epsilon_{\pm}^{~\nu_2*}(\hp)&=&
\half\left\{\Pi^{\mu_1\nu_1}(\hp)\Pi^{\mu_2\nu_2}(\hp)
+\Pi^{\mu_1\nu_2}(\hp)\Pi^{\mu_2\nu_1}(\hp)\right.\nn\\
&&~~~~~~~
\left.-\Pi^{\mu_1\mu_2}(\hp)\Pi^{\nu_1\nu_2}(\hp)\right\}.
 \eeqa

The polarization 'vectors' are not the Lorentz four vectors\footnote
{It comes from the fact that the little group is not semisimple for
massless fields. Translation in the little group isomorphic to
$ISO(2)$ generates the gradient term in the transformations. },
rather they transform as \beqa \label{trpol}
\epsilon_{\pm}^\mu(p)&=& e^{\mp i\Theta(\Lambda, p)}
[D_\epsilon^s(\Lambda,p)]^\mu_{~\nu}\epsilon_{\pm}^\nu(\Lambda p),\nn\\
{[D_\epsilon^s(\Lambda,p)]}^\mu_{~\nu} &=& (\Lambda^{-1})^\mu_{~\nu}
-\fr{p^\mu}{|\bp |}(\Lambda^{-1})^0_{~\nu}. \eeqa For general spin
$s$, the polarization tensor can be written as \beqa
\epsilon^A_{\sigma}(p) = e^{\mp i\Theta(\Lambda, p)}
[D_\epsilon^s(\Lambda,p)]^A_{~B} ~\epsilon^B_{\sigma}(\Lambda p),
\eeqa up to a phase factor.

\section{Invariant $M_\star$ function for massless field}
\label{apc}

Let $P$ denotes a shorthand notation for the external lines,
$P\equiv(p_1,\ldots,p_n)$, and $K$ denotes a standard
momenta\footnote{ see \cite{Weinberg},\cite{Stapp}. }
$K\equiv(k_1,\ldots,k_n)$. There exists a unique Lorentz
transformation satisfying $P\equiv L_P K$. Then the relation between
the Lorentz group and the little group can be described symbolically
as: \vs{0.1cm}
\begin{center}
\mbox{\large \xymatrix{ K  \ar[d]_{L_P} & & & K \ar[lll]_{W(\Lambda,
P)}\ar[d]^{L_{\Lambda P}}
\\
P \ar[rrr]_{\Lambda} & & & \Lambda P&, } }
\end{center}
\vs{0.1cm} where $W(\Lambda,P)$ is the Wigner transformation to
which a Lorentz transformation $\Lambda$ and the momenta $P$
correspond. Since the twist  do not change the group properties,
above relation also holds for the twisted symmetry group. The
$S_\star$ matrix elements transform as \beqa \label{St0} S_\star[P]
&=& D_\theta^s[W(\Lambda,P)]~S_\star[\Lambda P]\nn\\
&=& D_\theta^s[L^{~-1}_{P} \cdot\Lambda^{-1}\cdot L_{\Lambda
P}]~S_\star[\Lambda P] \eeqa where the indices for the external
lines are suppressed. From the golden rule and (\ref{taad}), the
explicit transformation formular for $S_\star[P]$ can be obtained:
\beqa \label{stl} S_\star[P] =\fr{N(\Lambda P)}{N(P)}~e^{\pm i
s\Theta(\Lambda, P)} ~S_\star[\Lambda P], \eeqa where $N$ denotes
the corresponding normalization factor.

If one defines $M_\star[P]$ as
$S_\star[P]=D_\theta^s[L^{~-1}_P]~M_\star[P]$, one can show that
$M_\star[P]$ transform as \beqa \label{covM} M_\star[P]=
D_\theta^s[\Lambda^{-1}]~M_\star[\Lambda P], \eeqa i.e., it is twist
invariant.  We will call it as an invariant $M_\star$ function as in
\cite{Stapp,BMW}.

Let us find out the invariant $M_\star$ function for massless
fields. First, we define the quanties for the standard momenta $K$.
The choice for $M_\star[K]$ is \beqa
M_\star^A[K]=N(K)~\epsilon^A[K]~S_\star[K]. \eeqa If we define
$M_\star[P]$ as \beqa
M_\star^A[P]=[D_\theta^s(L_P)]^A_{~B}~M_\star^B[K], \eeqa then one
can easily show that $M_\star^A[P]$ is twist invariant.

By using the explicit form of $D_\epsilon^s$ in (\ref{trpol}) and
$k_{b_j}M_\star^{b_1\cdots b_s}[K]=0$, one obtains the relation
\beqa \label{eM} \epsilon^{C}[P]^*~(M_\star)_C[P] &=& \left( e^{\pm
i s\Theta(L_P^{-1}, P)} [D^\epsilon_\theta(L_P^{-1})]^C_{~~A}
~\epsilon^A[K]^* \right) \left(
[D_\theta(L_P)]_C^{~~B}~(M_\star)_B[K] \right)
\nn\\
&=& e^{\pm i s\Theta(L_P^{-1}, P)}~\epsilon^A[K]^*
~[D_\theta(L_P^{-1})\cdot D^\epsilon_\theta(L_P^{-1})]^B_{~~A}~(M_\star)_B[K]\nn\\
&=&
e^{\pm i s\Theta(L_P^{-1}, P)}~\epsilon^{A}[P]^*~(M_\star)_A[P]\nn\\
&=&e^{\pm i s\Theta(L_P^{-1}, P)}N(K)~ S_\star[K], \eeqa where we
used the relation \beqa [D_\theta(L_P^{-1})\cdot
D^\epsilon_\theta(L_P^{-1})]^B_{~~A}~(M_\star)_B[K] &\equiv&
[D(L_P^{-1})\cdot D_\epsilon(L_P^{-1})]^B_{~~A}~(M_\star)_B[K]\nn\\
&=& \left(\delta^{b_1}_{~a_1}-
\fr{k^{b_1}}{|\bp|}(L_P)^0_{~a_1}\right) \cdots
\left(\delta^{b_s}_{~a_s}- \fr{k^{b_s}}{|\bp|}(L_P)^0_{~a_s}\right)
\times
(M_\star)_{b_1\cdots b_s}[K]\nn\\
&=& (M_\star)_{a_1\cdots a_s}[K]. \eeqa By setting $\Lambda =
L_P^{-1}$ in (\ref{stl}), one can see that (\ref{eM}) is
$S_\star[P]$. Thus, the desired form of $S_\star[P]$ is \beqa
S_\star[P]=N(P)~\epsilon^*_A[P]~M^A_\star[P]. \eeqa

\end{appendix}


\begin{thebibliography}{99}

\bibitem{Doplicher}
S. Doplicher, K. Fredenhagen, and J. E. Roberts, Comm. Math. Phys.
\textbf{172}, 187 (1995).

\bibitem{Weyl}
H. J. Groenewold, Physica \textbf{12}, 405 (1946); J. E. Moyal,
Proc. Cambridge Phil. Soc. \textbf{45}, 99 (1949); H. Weyl,
\textit{Quantum mechanics and group theory}, Z. Phys. \textbf{46}, 1
(1927).

\bibitem{chaichian}
M. Chaichian, P. P. Kulish, K. Nishijima, and A. Tureanu, Phys.
Lett. \textbf{B 604}, 98(2004).

\bibitem{Oeckl}
R. Oeckl, Nucl. Phys. \textbf{B 581}, 559 (2000).

\bibitem{Wess}
J. Wess, hep-th/0408080.

\bibitem{MajidRuegg}
S. Majid, H. Ruegg, Phys. Lett. \textbf{B 334}, 348 (1994).

\bibitem{Lukierski}
J. Lukierski, A. Nowicki, H. Ruegg and V. N. Tolstory, Phys. Lett.
\textbf{B 264}, 331 (1991).

\bibitem{chaichianprl}
M. Chaichian, P. Presnajder, and  A. Tureanu, Phys. Rev. Lett.
\textbf{94}, 151602 (2005).

\bibitem{Bala}
A. P. Balachandran, G. Mangano, A. Pinzul and S. Vaidya, Int. J.
Mod. Phys. \textbf{A 21}, 3111 (2006).

\bibitem{Bu}
J. G. Bu, H. C. Kim, Y. Lee, C. H. Vac, and J. H. Yee, Phys. Rev.
\textbf{D 73}, 125001 (2006).

\bibitem{Weinberg}
S. Weinberg, Phys. Rev. \textbf{135}, B1049 (1964).

\bibitem{Weinberg1}
S. Weinberg, Phys. Rev. \textbf{133}, B1318 (1964).

\bibitem{Weinberg2}
S. Weinberg, Phys. Rev. \textbf{134}, B882 (1964).

\bibitem{Bonora}
L. Bonora, M. Schnabl, M. M. Sheikh-Jabbari, A. Tomasiello, Nucl.
Phys. \textbf{B 589}, 461 (2000).

\bibitem{WessGa}
J. Madore, S. Schraml, P. Schupp, J. Wess, Eur. Phys. J.
\textbf{C16}, 161 (2000).

\bibitem{Chaichiangauge}
M. Chaichian, P. Presnajder, M. M. Sheikh-Jabbari, A. Tureanu, Phys.
Lett. \textbf{B 526}, 132 (2002).

\bibitem{Fatoll}
A. H. Fatollahi, H. Mohammadzadeh,
 Eur. Phys. J. \textbf{C 36} 113 (2004).

\bibitem{Rivelles}
V. O. Rivelles, Phys. Lett. \textbf{B 558}, 191 (2003).

\bibitem{Banados}
M. Banados, O. Chandia, N. Grandi, F. A. Schaposnik, G. A. Silva,
Phys. Rev. \textbf{D64}, 084012 (2001).

\bibitem{WessGr}
P. Aschieri, C. Blohmann, M. Dimitrijevic, F. Meyer, P. Schupp, J.
Wess, Class. Quant. Grav. \textbf{22}, 3511 (2005).

\bibitem{Yang}
R. Banerjee, H. S. Yang, Nucl. Phys. \textbf{B708}, 434 (2005).

\bibitem{Ellis}
 J. Ellis, N. E. Mavromatos, D. V. Nanopoulos, A. S. Sakharov,
 Int. J. Mod. Phys. \textbf{A19}, 4413 (2004).

\bibitem{Adunas}
G. Z. Adunas, E. Rodriguez-Milla, D. V. Ahluwalia, Gen. Rel. Grav.
\textbf{33}, 183 (2001).

\bibitem{Obukhov}
Y. N. Obukhov, Phys. Rev. Lett. \textbf{86}, 192 (2001).


\bibitem{Majid}
S. Majid, \textit{Foundations of Quantum Group Theory}, Cambridge
University Press, (1995).


\bibitem{Filk}
T. Filk, Phys. Lett. \textbf{B 376}, 53 (1996).

\bibitem{Joung}
E. Joung, J. Mourad, hep-th/0703245.

\bibitem{WeinbergB}
S. Weinberg, \textit{The Quantum Theory of Fields}, Vol.I,
 Cambridge University Press, (1996).


\bibitem{Stapp}
H. Stapp, Phys. Rev. \textbf{125}, 2139 (1962).


\bibitem{BMW}
A. O. Barut, I. Muzinich, D. N. Williams, Phys. Rev. \textbf{130},
442 (1963).


\end{thebibliography}
\end{document}